\documentclass[twocolumn,showpacs,preprintnumbers,amsmath,amssymb]{revtex4}

\usepackage{graphicx}
\usepackage{dcolumn}
\usepackage{bm}
\usepackage{epsfig}
\usepackage{graphics}
\usepackage{graphicx}
\usepackage{mathtext}

\begin{document}

\title{
Scale-free network topology and multifractality in weighted planar stochastic lattice
}%

\author{M. K. Hassan$^{1}$, M. Z. Hassan$^{2}$ and N. I. Pavel$^{1}$
}%
\date{\today}%

\affiliation{
$1$  Theoretical Physics Group, Department of Physics, University of Dhaka, Dhaka 1000, Bangladesh \\
$2$ Institute of Computer Science, Bangladesh Atomic Energy Commission, Dhaka 1000, Bangladesh 
}

\begin{abstract}%

We propose a weighted planar stochastic lattice (WPSL) formed by the random sequential partition of a plane 
into contiguous and non-overlapping blocks and find that it evolves following several non-trivial conservation laws, namely 
$\sum_i^N x_i^{n-1} y_i^{4/n-1}$ is independent of time $\forall \ n$, where $x_i$ and $y_i$ are the length and width of the $i$th block. 
Its dual on the other hand, obtained by replacing each block with a node at its center and common border between blocks with an edge joining 
the two vertices, emerges as a network with a power-law degree distribution $P(k)\sim k^{-\gamma}$ where $\gamma=5.66$ revealing scale-free 
coordination number disorder since $P(k)$ also describes the fraction of blocks having $k$ neighbours. To quantify the size disorder, 
we show that if the $i$th block is populated with $p_i\sim x_i^3$ then its distribution in the WPSL exhibits multifractality.

\end{abstract}


\maketitle

\section{Introduction}

Planar cellular structures formed by tessellation, tiling, or subdivision 
of plane into contiguous and nonoverlapping cells have always generated interest among scientists
in general and physicist in particular because cellular structures are ubiquitous in nature. 
Examples include acicular texture in martensite growth, tessellated pavement on ocean shores,
agricultural land division according to ownership, 
grain texture in polycrystals, cell texture in biology, 
soap froths and so on \cite{ref.martensite, ref.polycrystal,ref.biocell,ref.soapfroths}.
For instance, Voronoi lattice formed by partitioning of a plane into convex polygons or Apollonian packing 
generated by tiling a plane into contiguous and non-overlapping disks has found widespread applications 
\cite{ref.model,ref.apollonian}.
There are some theoretical models, both random and deterministic, developed either to directly mimic these structures or
to serve as a tool on which one can study various physical problems  \cite{ref.martensite}. 
However, cells in most of the existing structures do not have sides 
which share with more than one side of other cells. In reality, the 
sides of a cell can share a part or a whole side of another cell. As a result, the number of neighbours of a cell can be higher
than the number of sides it has. 
Moreover, cellular structure may emerge through evolution where cells can be of different sizes and have different 
number of neighbours since nature favours these properties as a matter of rule rather than exception. 
A lattice with such properties can be of great interest as it can mimic disordered medium
on which one can study percolation or random walk like problems.

\begin{figure}
\includegraphics[width=8.5cm,height=8.0cm,clip=true]{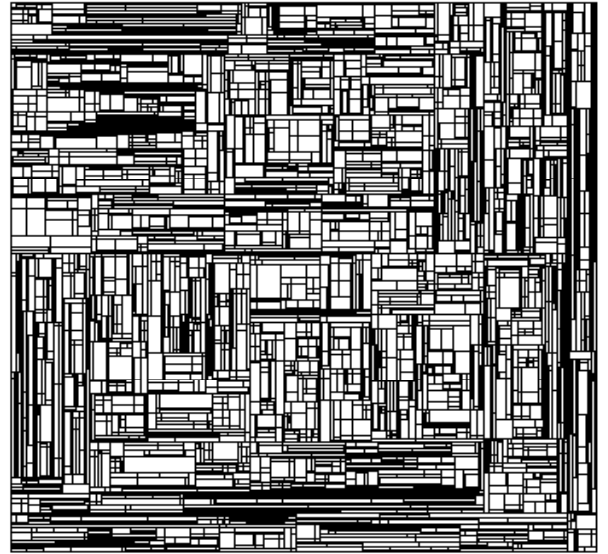}
\caption{A snapshot of the weighted planar stochastic lattice.
}
\label{fig1}
\end{figure}%

In this article, we propose a weighted planar
stochastic lattice (WPSL) as a space-filling cellular structure where annealed coordination number disorder and size disorder are 
introduced in a natural way. The definition of the model is trivially simple. It starts with an initiator, say a
square of unit area, and a generator that divides it randomly into four blocks. The generator thereafter is sequentially applied
over and over again to only one of the available blocks picked preferentially with respect to their areas. 
It results in the partitioning of the square into ever smaller mutually exclusive rectangular 
blocks. A snapshot of the WPSL at late stage (figure 1) provides an awe-inspiring perspective on 
the emergence of an intriguing and rich pattern of blocks. 
We intend to investigate its topological and geometrical properties in an attempt to find some order in this seemingly disordered 
lattice.
If the blocks of the WPSL are regarded as isolated fragments then the model can also describe the fragmentation of a $2$D 
object by random sequential nucleation of seeds from which two orthogonal cracks parallel to the
sides of the parent object are grown until intercepted by existing cracks \cite{ref.hassan,ref.krapivsky}.
In reality, fragments produced in fracture of solids by
propagation of interacting cracks is a formidable mathematical problem.
The model in question can, however, be considered as the minimum model which should be capable of capturing the essential 
features of the underlying mechanism. The WPSL can also describe the martensite formation 
as we find its definition remarkably similar to the model 
proposed by Rao {\it et al.} which is also reflected in the similarity between the figure 1  
and figure 2 of \cite{ref.martensite,ref.martensite_1}. Yet another application, perhaps a little exotic, is the 
random search tree problem in computer science \cite{ref.majumdar}.

Searching for an order in the disorder is always an attractive proposition in physics. To this end, we invoke the concept of complex network 
topology to quantify the coordination number disorder and the idea of multifractality to quantify the size disorder of the blocks in WPSL. 
It is interesting to note that the dual of the WPSL (DWPSL) obtained by replacing each block with a node at its center and common border 
between blocks with an edge joining the two corresponding vertices emerges as a network. 
The area of the respective blocks is assigned to the corresponding nodes to characterize them as their fitness parameter. Nodes
in the DWPSL, therefore, are characterized by their respective fitness parameter and the corresponding degree $k$ defined as the number of links a node has. 
For a decade, there has been a surge of interest in finding the degree distribution $P(k)$ 
triggered by the work of A.-L. Barabasi and his co-workers who have revolutionized the notion of 
the network theory by recognizing the fact that real networks are not static rather  
grow by addition of new nodes establishing links preferentially, known as the preferential attachment (PA) rule, 
to the nodes that are already well connected \cite{ref.barabasi}.  Incorporating both the ingredients, growth and the PA rule, 
Barabasi and Albert (BA) presented a simple theoretical model and showed that such network 
self-organizes into a power-law degree distribution $P(k)\sim k^{-\gamma}$
with $\gamma=3$  \cite{ref.review}. The phenomenal success of the BA model lies in the fact that it can capture, at least qualitatively, the key 
features of many real life networks.
Interestingly, we find that the DWPSL has all the ingredients of the BA model and its degree distribution $P(k)$ follows heavy-tailed
power-law but with exponent $\gamma=5.66$ revealing that the coordination number of the WPSL is scale-free in character.

In addition to characterizing the blocks of the WPSL by the coordination number $k$, they can also be characterized by their 
respective length $x$ and width $y$. 
We then find that the dynamics of the WPSL is governed by infinitely many conservation laws, namely the quantity 
$\sum_ix_i^{n-1}y_i^{4/n-1}$ remains independent of time $\forall$ $n$ where blocks are labbeled by index $i=1,2,...,N$. 
For instance, total area is one obvious conserved quantity obtained by setting $n=2$, sum of the cubic power of the length (or width) of 
all the existing blocks $\sum_i^N x_i^3$ is a non-trivial conserved quantity  obtained by setting $n=1$ (or $n=4$).
Interestingly, we find that when the $i$th block is populated with the fraction of the measure $p_i\sim x_i^3$  
then the distribution of the population in the WPSL emerges as multifractal
indicating further development towards gaining deeper insight into the complex nature of the WPSL we proposed. 
Multifractal analysis was initially proposed to treat
turbulence but later successfully applied in a wide range of exciting field of research \cite{ref.mandelbrot}.
Recently though it has received a renewed interest as it has been found that the wild fluctuations of the wave functions in the
vicinity of the Anderson and the quantum Hall 
transition can be best quantified by using a multifractal analysis \cite{ref.anderson}. 

The organization of this paper is as follows. In section 2, we give its exact algorithm of the model. In section 3 various structural topological 
properties of the WPSL and its dual are discussed in order to quantify the annealed coordination number disorder. In section 4 we discuss the 
geometric properties of the WPSL in an attempt to quantify the annealed size disorder. Finally, section 5 gives a short summary of our results.

\begin{figure}
\includegraphics[width=8.50cm,height=4.5cm,clip=true]{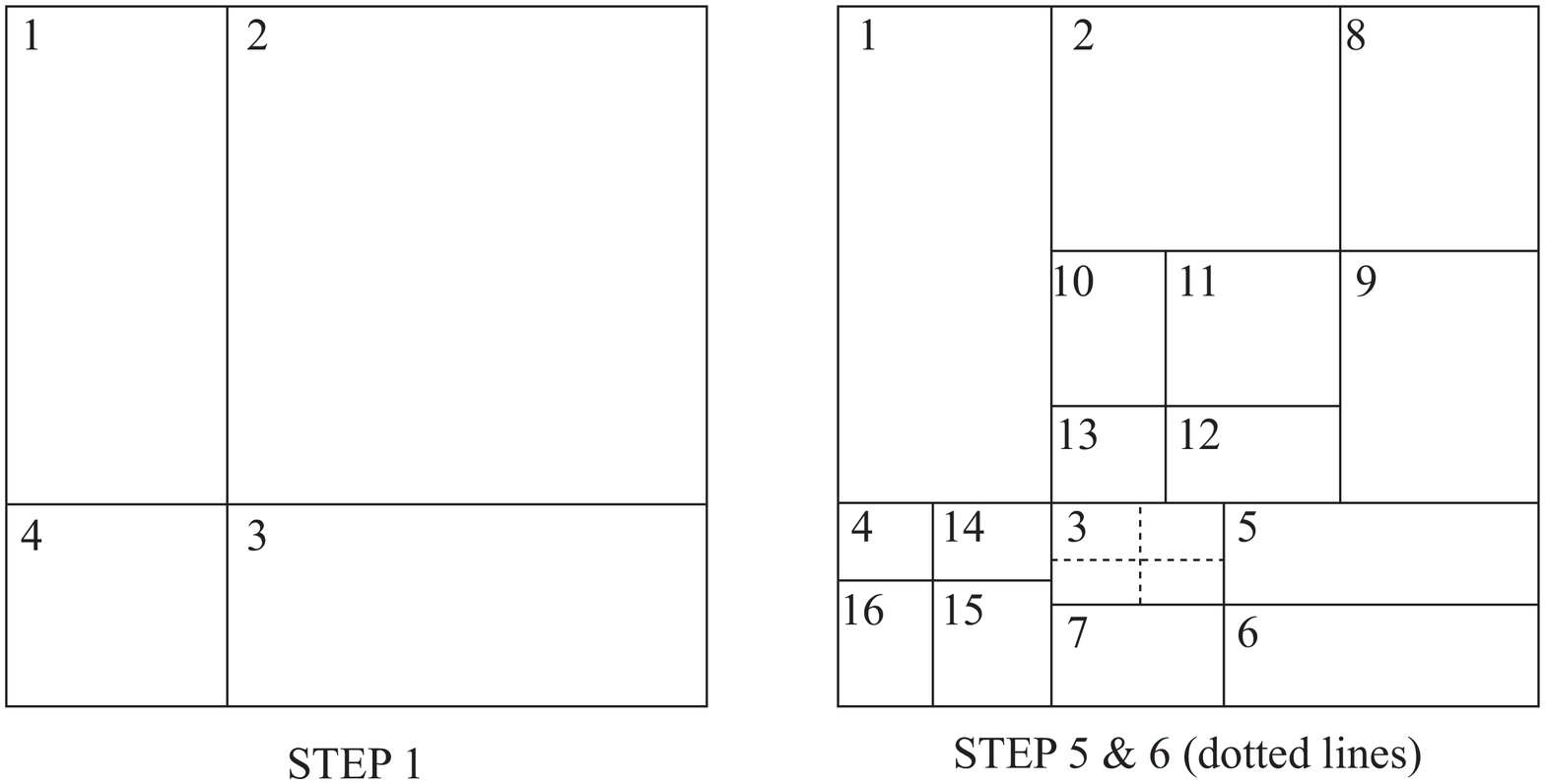}
\caption{Schematic illustration of the first few steps of the algorithm.
}
\label{fig2}
\end{figure}%

\section{Algorithm of the model}

Perhaps an exact algorithm can provide a better description of the model than the mere definition.
In step one, the generator divides the initiator,
say a square of unit area, randomly into four smaller blocks. We then label the four newly created blocks 
by their respective areas $a_1, a_2, a_3$ and $a_4$ in a clockwise fashion starting from the upper left block (see figure 2). In each step thereafter
only one block is picked preferentially with respect to their respective area (which we also refer as the fitness parameter) 
and then it is divided randomly into four blocks. In general, the $j$th step of the algorithm can be described as follows.
(i) Subdivide the interval $[0,1]$ into $(3j-2)$ subintervals of size $[0,a_1]$, $[a_1, a_1+a_2]$,$\  ...$, 
$[\sum_{i=1}^{3j-3} a_i,1]$ each of which represents the blocks labelled by their areas $a_1,a_2,...,a_{(3j-2)}$ respectively.
(ii) Generate a random number $R$ from the interval $[0,1]$ and find which of the $(3i-2)$ sub-interval 
contains this $R$. The corresponding block it represents, say the $p$th block of area $a_p$, is picked.  
(iii) Calculate the length $x_p$ and the width $y_p$ of this block and keep note of the coordinate of the
lower-left corner of the $p$th block, say it is $(x_{low}, y_{low})$.
(iv) Generate two random numbers $x_R$ and $y_R$ from $[0,x_p]$ and $[0,y_p]$ respectively and hence
the point $(x_{R}+x_{low},y_{R}+y_{low})$ mimics a random point chosen in the block $p$.
(v) Draw two perpendicular lines through the point $(x_{R}+x_{low},y_{R}+y_{low})$ 
parallel to the sides of the $p$th block in order to divide it into four smaller blocks. The label $a_p$ is now redundant and hence
it can be reused.
(vi) Label the four newly created blocks according to their areas $a_p$, $a_{(3j-1)}$, $a_{3j}$ and $a_{(3j+1)}$ respectively 
in a clockwise fashion starting from the upper left corner.
(vii) Increase time by one unit and repeat the steps (i) - (vi) {\it ad infinitum}.

\begin{figure}
\includegraphics[width=8.50cm,height=5.5cm,clip=true]{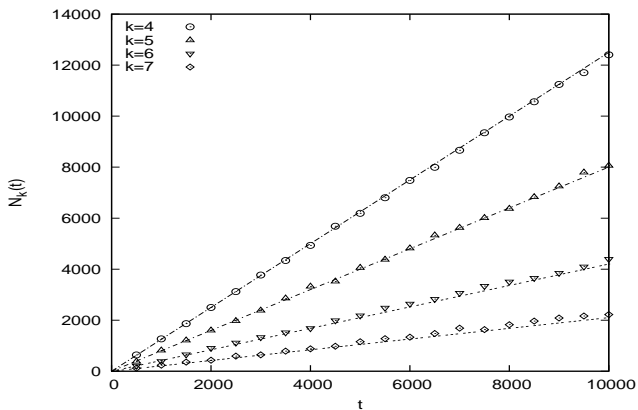}
\caption{Shown is the growth of the number of blocks $N_k(t)$ with exactly $k=4,5,6,7$
neighbours as a function of time.
}
\label{fig3}
\end{figure}%

\section{Topological properties of WPSL}

We first focus our analysis on the blocks of the WPSL and their coordination numbers. Note that for square lattice, 
the deterministic counterpart of the WPSL, the coordination number is a constant equal to $4$. However, the coordination number in the WPSL 
is neither a constant nor has a typical mean value, rather the coordination number that each 
block assumes in the WPSL is random. Moreover, it is allowed to evolve with time 
and hence the coordination number disorder in the WPSL can be regarded as of annealed type. 
Defining each step of the algorithm as one time unit and imposing periodic boundary condition in the simulation, we 
find that the number of blocks $N_k(t)$ which have coordination number $k$ (or $k$ neighbours) continue to
grow linearly with time $N_k(t)=m_k t$ (see figure 3). On the other hand, the number of total blocks $N(t)=\sum_k N_k(t)$ at time $t$
in the lattice also grow linearly with time $N(t)=1+3t$ which we can write $N(t)\sim 3t$ in the asymptotic regime. 
The ratio of the two quantities $\rho_k(t)=N_k(t)/N(t)$ that
describes the fraction of the total blocks which have coordination number $k$ is
$\rho_k(t)=m_k/3$. It implies that $\rho_k(t)$ becomes a global property  since we find it independent of time and 
size of the lattice. We now take WPSL of fixed size or time and look into its structural properties. 
For instance, we want to find out what fraction of the total blocks of the WPSL of a given size has coordination number $k$. For this
we collect data for $\rho_t$ as a function of $k$ and obtain the coordination number distribution function $\rho_t(k)$ where subscript
$t$ indicates fixed time. 
Interestingly, the same WPSL can be interpreted as a network if the blocks of the lattice are regarded as 
nodes and the common borders between blocks as their links which is topologically identical to the dual of the WPSL (DWPSL).

\begin{figure}
\includegraphics[width=8.50cm,height=5.5cm,clip=true]{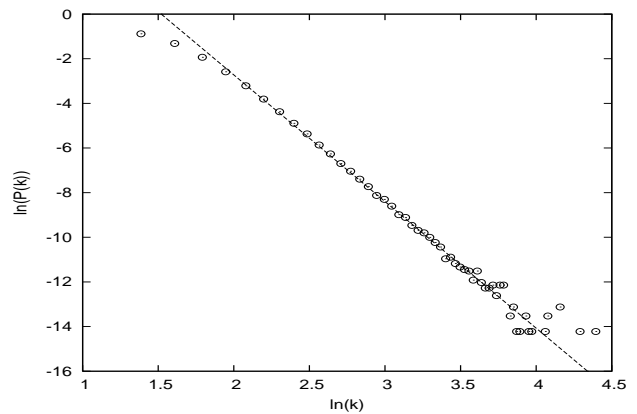}
\caption{Degree distribution $P(k)$ for the DWPSL network where data points represent average of $50$ independent realizations.
The line have slope exactly equal to $5.66$ revealing power-law degree distribution with exponent $\gamma=5.66$.
}
\label{fig4}
\end{figure}%

\begin{figure}
\includegraphics[width=8.50cm,height=5.5cm,clip=true]{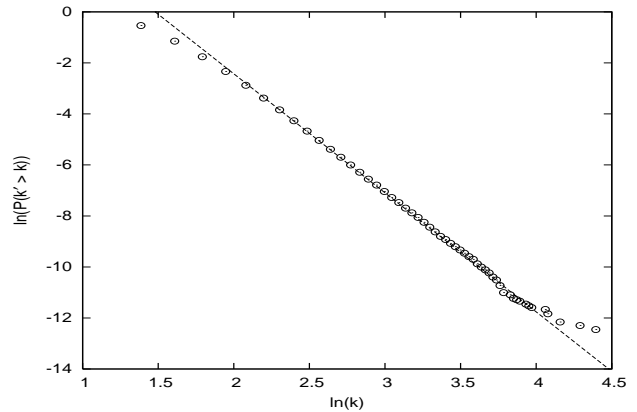}
\caption{Cumulative degree distribution $P(k^\prime>k)$ is shown using using the same data as of figure 4.
The dotted line with slope equal to $\gamma-1=4.66$ is drawn to guide our eyes.
}
\label{fig5}
\end{figure}%

In fact, the data for the degree distribution $P(k)$ of the 
DWPSL network is exactly the same as the coordination number distribution function $\rho_t(k)$ of the WPSL i.e., $\rho_t(k)\equiv P(k)$. 
The plot of $\ln P(k)$ vs $\ln (k)$ in figure 4 using data obtained after ensemble average over $50$ independent realizations 
clearly suggests that fitting a straight line to data is possible. This implies that the degree distribution
decays obeying power-law 
\begin{equation}
\label{degreedistribution}
P(k)\sim k^{-\gamma}. 
\end{equation}
However, note that figure 4 has heavy or fat-tail,
benchmark of the scale-free network, which represents highly connected {\it hub} nodes. The presence of 
messy tail-end complicates the process of fitting the data
into power-law forms, estimating its exponent $\gamma$, and identifying the range over which 
power-law holds. One way of reducing the noise at the tail-end of the degree distribution is to plot   
cumulative distribution $P(k^\prime>k)$. We therefore plot  $\ln (P(k^\prime >k))$ vs $\ln (k)$ in figure 5 
using the same data of figure 4 
and find that the heavy tail smooths out naturally where no data is obscured. 
The straight line fit of figure 5 has a slope $\gamma-1=4.66$ which 
indicates that the degree distribution (figure 4) decays following power-law with exponent $\gamma=5.66$. We find it worthwhile 
to mention as a passing note that the mean coordination number $\sum_k kP(k)$ reaches asymptotically to a constant value equal to $5.333$.

We thus find that in the large-size limit 
the WPSL develops some order in the sense that 
its annealed coordination number disorder or the degree distribution of its dual is scale-free in character. 
This is in sharp contrast to the quenched coordination number disorder found in the Voronoi lattice where 
it is almost impossible to find cells which have
significantly higher or fewer neighbours than the mean value $k=6$ \cite{ref.vd}. In fact, it has been shown that 
the degree distribution of the dual of the Voronoi lattice
is Gaussian. The square lattice, on the other hand, is self-dual and its degree distribution is $P(k)=\delta(k-4)$.
Further, it is  interesting to point out that the exponent $\gamma=5.66$ is significantly higher than 
usually found in most real-life network which is typically $2<\gamma\leq 3$. This suggests that in addition to the PA rule the network
in question has to obey some constraints. For instance, nodes in the WPSL are spatially embedded in Euclidean space,
links gained by the incoming nodes are constrained by the spatial location and the fitting parameter of the nodes.  
Owing to its unique dynamics this was not unexpected.
Perhaps, it is noteworthy to mention that the the degree distribution  of the electric power grid, whose nodes like WPSL are also embedded in
the spatial position, is shown to exhibit power-law but with exponent $\gamma_{{\rm power}}=4$  \cite{ref.barabasi}.

The power-law degree distribution $P(k)$ has been found
in many seemingly unrelated real life networks. It implies that there must exists some common 
underlying mechanisms for which disparate systems behave in such a 
remarkably similar fashion \cite{ref.review}. Barabasi and Albert argued that the growth and the PA rule are the 
main essence behind the emergence of such power-law. Indeed, the DWPSL network too grows with time but in sharp contrast to the BA model 
where network grows by addition of one single node with $m$ edges per unit time,
the DWPSL network grows by addition of a group of three nodes which are already linked by two edges. 
It also differs in the way incoming nodes establish links with the existing nodes.

To understand the growth mechanism of the DWPSL network let us look into the $j$th step of the algorithm. First a node, say it is labeled as $a_p$, is picked from the $(3j-2)$ 
nodes preferentially with respect to the fitness parameter of a node (i.e., according to their respective areas). Secondly, connect the node $a_p$
with two new nodes $(3j-1)$ and $(3j+1)$ in order to establish their links with the existing network.
At the same time, at least two or more links of $a_p$ with other nodes are removed (though the exact number depends on the number of neighbours $a_p$ already 
has) in favour of linking them among the three incoming nodes in a self-organised fashion. 
In the process, the degree $kj_p$ of the node $a_p$ will either decrease (may turn into a node with marginally low 
degree in the case it is highly connected node) or at best remain the same but will never increase. 
It, therefore, may appear that the PA rule is not followed here. 
A closer look into the dynamics, however, reveals otherwise. It is interesting to note that an existing nodes during the process gain links
only if one of its neighbour is picked not itself. It implies that the higher the links (or degree) a node has, the higher 
its chance of gaining more links since they can be reached in a larger number of ways. It essentially embodies the intuitive idea of PA rule.
Therefore, the DWPSL network can be seen to follow preferential attachment rule but in disguise.  

\section{Geometric properties of WPSL}

We again focus on the blocks of the WPSL but this time we characterize them by their length $x$ and width $y$ instead of the number of neighbours
they have. Then  
the evolution of their distribution function $C(x,y;t)$ can be described by the following kinetic equation
\cite{ref.hassan,ref.krapivsky}
\begin{eqnarray}
{{\partial C(x,y;t)}\over{\partial t}}& =& -xyC(x,y;t)+ \\ \nonumber & & 
4\int_x^\infty\int_y^\infty C(x_1,y_1;t)dx_1dy_1.
\end{eqnarray}
Incorporating the $2$-tuple Millen transform 
\begin{equation}
M(m,n;t)=\int_0^\infty\int_0^\infty x^{m-1}y^{n-1}C(x,y;t)dxdy,
\end{equation}
in the kinetic equation yields
\begin{equation}
\label{eq:momenteq}
{{dM(m,n;t)}\over{dt}}=\Big ( {{4}\over{mn}}-1\Big )M(m+1,n+1;t).
\end{equation}
Iterating it to get all the derivatives of $M(m,n;t)$ and
then substituting them into the Taylor series expansion of $M(m,n;t)$ about $t=0$ one
can write its solution in terms of generalized hypergeometric function \cite{ref.hypergeometric}
\begin{equation}
M(m,n;t)=  ~_2F_2\Big (a_+,a_-;m,n;-t\Big ),
\end{equation}
where $M(m,n;t)=M(n,m;t)$ for symmetry reason and 
\begin{equation}
a_{\pm} = {{m+n}\over{2}} \pm \Big [ \Big ({{m-n}\over{2}}\Big )^2+4 \Big ]^{{{1}\over{2}}}.
\end{equation}  
One can immediately see that (i) $M(1,1;t)=1+3t$ is the total
number of blocks $N(t)$ and (ii) $M(2,2;t)=1$ is the total area of all the blocks
which is obviously a conserved quantity. The behaviour of $M(m,n;t)$ in the long time limit is
\begin{equation}
\label{eq:aminus}
M(m,n;t)\sim t^{-a_-}.
\end{equation}
It implies that the system is in fact governed by several conservation laws namely, 
$M(n,4/n;t)=\sum_i^N x_i^{n-1} y_i^{{{4}\over{n}}-1}$ are independent of time 
for all $n$ which has been confirmed by numerical simulation (see figure 6).

\begin{figure}
\includegraphics[width=8.50cm,height=5.5cm,clip=true]{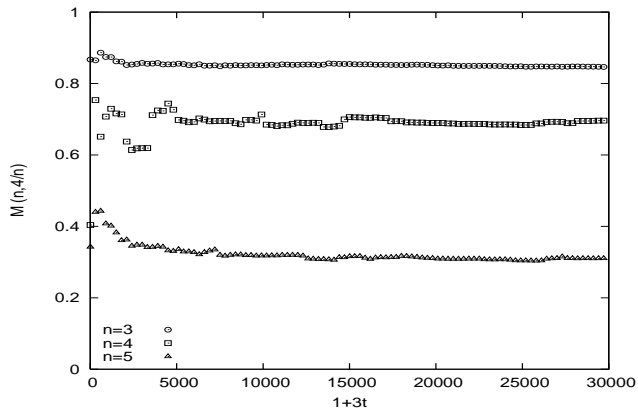}
\caption{The plots of $\sum_i^N x_i^{n-1}y_i^{4/n-1}$ vs $N$ for $n=3,4,5$ are drawn using data collected from
one realization.
}
\label{fig6}
\end{figure}%

We now focus on the distribution function $n(x,t)=\int_0^\infty C(x,y,t)dy$ that describes
the concentration of blocks of length $x$ at time $t$ regardless of the size of their widths $y$. Then the $q$th moment of $n(x,t)$ is
defined as 
\begin{equation}
M_q(t)=\int_0^\infty x^q n(x,t)dx.
\end{equation}
Appreciating the fact that $M_q(t)=M(q+1,1;t)$ and using equation (\ref{eq:aminus}) one can immediately write that 
\begin{equation}
\label{eq:qmoment}
M_q(t)\sim t^{\{\sqrt{q^2+16}-(q+2)\}/2}.
\end{equation}
Note that had we focus on $n(y,t)$ instead of $n(x,t)$ we would have exactly the same result since their $q$th moments are identical,
$M(q+1,1;t)=M(1,q+1;t)$, due to symmetry reason. 
We find that the quantity $M_3(t)$ and hence $\sum_i^N x_i^3$ or $\sum_i^N x_i^3$ is a conserved quantity. 
However, yet another interesting
fact is that although $M_3(t)$ remains a constant against time in every independent realization, their exact numerical value fluctuates
sample to sample (see figure 7). It clearly indicates lack of self-averaging or wild fluctuation. 
Nonetheless, during each realization we can use $M_3(t)$ as a measure to populate the $i$th block with the fraction of the 
total population $p_i=x_i^3/\sum_ix_i^3$.
The corresponding "partition function" then is 
\begin{equation}
Z_q(t)=\sum_ip_i^q\sim M_{3q}(t),
\end{equation}
whose solution can be written immediately from equation (\ref{eq:qmoment}) to give
\begin{equation}
Z_q(t)\sim t^{\{\sqrt{9q^2+16}-(3q+2)\}/2}.
\end{equation}

\begin{figure}
\includegraphics[width=8.50cm,height=5.5cm,clip=true]{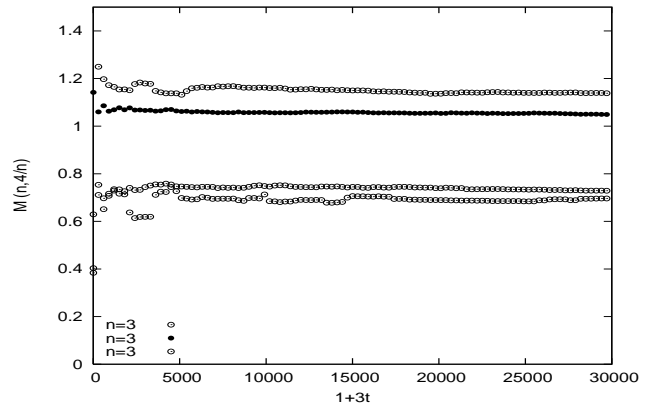}
\caption{The plots of $\sum_i^N x_i^3$ vs $N$ for four different realizations shows that 
the numerical value is different at every independent realization.
}
\label{fig7}
\end{figure}%

We find it instructive to express $Z_q$ in terms of square root of the mean area
$\delta(t)=\sqrt{M(2,2;t)/M(1,1;t)}\sim t^{-1/2}$ as it gives the weighted number
of squares $N(q,\delta)$ needed to cover the measure which scales as
\begin{equation}
\label{weightednumber}
N(q,\delta)\sim \delta^{-\tau(q)},
\end{equation}
where the mass exponent 
\begin{equation}
\label{massexponent}
\tau(q)=\sqrt{9q^2+16}-(3q+2).
\end{equation} 
The non-linear nature of $\tau(q)$ suggests
that an infinite hierarchy of
exponents is required to specify how the moments of the probabilities $\{p\}$s 
scales with $\delta$. Note that $\tau(0)=2$ is the 
Hausdorff-Besicovitch (H-D) dimension  of the WPSL since the bare number $N(q=0,\delta)\sim \delta^{-\tau(0)}$
and $\tau(1)=0$ follows from the conservation laws (or normalization of the probabilities $\sum p_i=1$).

\begin{figure}
\includegraphics[width=8.50cm,height=5.5cm,clip=true]{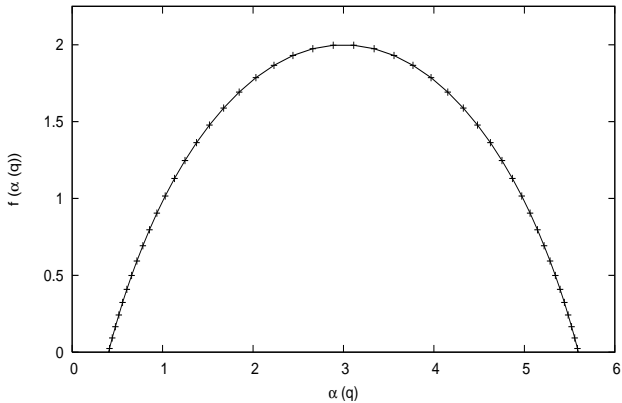}
\caption{The $f(\alpha)$ spectrum.
}
\label{fig8}
\end{figure}%

We now perform the Legendre transformation of $\tau(q)$ by using the Lipschitz-H\"{o}lder exponent 
\begin{equation}
\alpha=-{{d\tau(q)}\over{dq}},
\end{equation} 
as an independent variable to obtain the new function 
\begin{equation}
f(\alpha)=q\alpha+\tau(q).
\end{equation}
Substituting it in equation (\ref{weightednumber}) we find that 
\begin{equation}
N(q,\delta) \sim \delta^{q\alpha-f(\alpha)},
\end{equation}
since it is the dominant value of the integral  
\begin{equation}
N(\alpha,\delta) \sim \int \rho(\alpha)d\alpha \delta^{-f(\alpha)}\delta^{q\alpha},
\end{equation}
obtained by the extremal conditions. We can infer from this that the number of squares $dN(\alpha,\delta)$ needed to 
cover the PSL with 
$\alpha$ in the range $\alpha$ to $\alpha+d\alpha$ scales as
\begin{equation}
\rho(\alpha)d\alpha\delta^{-f(\alpha)},
\end{equation}
where $\rho(\alpha)d\alpha$ is the number of times the subdivided regions indexed by $\alpha$ \cite{ref.multifractal_1}.
It implies that a spectrum of spatially intertwined fractal 
dimensions 
\begin{equation}
f(\alpha(q))={{16}\over{\sqrt{9q^2+16}}}-2,
\end{equation}
are needed to characterize the measure which is always concave in character (see figure 8). 
It implies that
the size disorder of the blocks are multifractal in character since the measure $\{p_\alpha\}$ is related
to size of the blocks.  
That is, the distribution of $\{p_\alpha\}$ in WPSL can
be subdivided into a union of fractal subsets each with fractal dimension $f(\alpha)\leq 2$ in which the 
measure $p_\alpha$ scales as $\delta^\alpha$.
Note that $f(\alpha)$ is always concave in character (see figure 8) with a single maximum at $q=0$ correspond to the dimension of the WPSL 
with empty blocks.

On the other hand, we find that the entropy $S(\delta)=-\sum_i p_i\ln p_i$ associated with the partition of the measure on the support (WPSL) by 
using the relation $\sum_i p_i^q\sim \delta^{-\tau(q)}$ in the definition of $S(\delta)$. Then  
a few steps of algebraic manipulation reveals that $S(\delta)$ 
exhibits scaling 
\begin{equation}
S(\delta)=\ln\delta^{-\alpha_1}
\end{equation} 
where the exponent $\alpha_1={{6}\over{5}}$ obtained from 
\begin{equation}
\alpha_q= -\left. d\tau(q)/dq \right |_{q}.
\end{equation} 
It is interesting to note that $\alpha_1$ is related to the generalized dimension $D_q$, also related to the R\'{e}nyi entropy 
$H_q(p)={{1}\over{q-1}}\ln \sum_i p_i^q$ in the 
information theory, given by
\begin{equation}
\label{dq}
D_q=\lim_{\delta\rightarrow 0} \Big [{{1}\over{q-1}}{{\ln \sum_i p_i^q}\over{\ln\delta}}\Big ]={{\tau(q)}\over{1-q}},
\end{equation}
which is often used in the multifractal formalism as it can also provide insightful interpretation. 
For instance, $D_0=\tau(0)$ is the dimension of the support, $D_1=\alpha_1$ is the R\'{e}nyi information dimension 
and $D_2$ is known as the correlation dimension \cite{ref.procaccia,ref.renyi_entropy}.

\section{Summary}

We have proposed and studied a weighted planar stochastic lattice (WPSL) which has annealed coordination number 
and the block size disorder.
We have shown that the coordination number disorder is scale-free in character 
since the degree distribution of its dual (DWPSL), which is topologically identical to the network obtained
by considering blocks of the WPSL as nodes and the common border between blocks as links, exhibits power-law.  
However, the novelty of this network is that it grows by addition of a group of 
already linked nodes which then establish links with the existing nodes
following PA rule though in disguise in the sense that existing nodes gain links only if one of their neighbour is picked not itself. 
Besides, we have shown that if the blocks of the WPSL are characterized by their respective length $x$ and width $y$
then we find $\sum_i^N x_i^{n-1} y_i^{4/n-1}$ remains a constant regardless of the size of the lattice. However, the numerical
values of the conserved quantities except $n=2$ varies from sample to sample revealing absence of self-averaging - an indication of wild fluctuation. 
We have shown that if the blocks are occupied with a fraction of the measure equal to cubic power 
of their respective length or width then its distribution on the WPSL is multifractal nature.  
Such multifractal lattice with scale-free coordination disorder can be of great interest as it has the potential to mimic 
disordered medium on which one can study various physical phenomena like percolation and random walk problems etc. 

NIP acknowledges support from the Bose Centre for Advanced Study and Research in Natural Sciences.

\end{document}